\documentclass[12pt, draftclsnofoot, onecolumn]{IEEEtran}
\usepackage{amsmath,theorem,cite}
\usepackage{amssymb}
\usepackage{graphicx}
\usepackage{color}
\usepackage[latin1]{inputenc}

\newtheorem{proposition}{Proposition}
\newtheorem{corollary}{Corollary}

\newtheorem{example}{Example}

\newtheorem{relationship}{Relationship}

\newcommand{\be}{\begin{equation}}
\newcommand{\ee}{\end{equation}}
\newcommand{\bea}{\begin{eqnarray}}
\newcommand{\eea}{\end{eqnarray}}
\newcommand{\beaa}{\begin{eqnarray*}}
\newcommand{\eeaa}{\end{eqnarray*}}

\def\E{{\mathbb E}}

\def\SNR    {\mbox{\scriptsize\sf SNR}}

\def\SNRsat    {\mbox{\scriptsize\sf SNR}_{\sf sat}}
\def\SIR    {\mbox{\scriptsize\sf SIR}}
\def\SINR   {\mbox{\scriptsize\sf SINR}}

\def\MMSE    {\mbox{\scriptsize\sf MMSE}}

\def\Psat {{P_{\sf sat}}}

\def\diag   {\mbox{\rm diag}}

\def \det       {{\rm det}}
\def \diag      {{\rm diag}}

\def \Tr        {\mathrm{Tr}}

\def\non{\nonumber\\}

\def\Idm{{\bf I}}

\def\Xm{{\bf X}}
\def\Ym{{\bf Y}}

\def\Gm{{\bf G}}

\def\Sm{{\bf S}}

\def\Hm{{\bf H}}

\def\AbvGT #1#2{\lower2pt\vbox{\baselineskip0pt \lineskip-.5pt%
         \halign{$#1 ##$\cr #2\crcr >\cr}}}

\def\fd {f_{\rm \scriptscriptstyle D}}
\def\Tc  {T_{\rm c}}

\def\Bc  {B_{\rm c}}

\def\complex{\mathop{\raise .45ex\hbox{${\bf\scriptstyle{|}}$}
      \kern -0.40em {\rm \textstyle{C}}}\nolimits}
\def\hilbert{\mathop{\raise .21ex\hbox{$\bigcirc$}}\kern -1.005em
{\rm\textstyle{H}}} 

\def\squarebox#1{\hbox to #1{\hfill\vbox to #1{\vfill}}}

\title{\bf Fundamental Limits of Cooperation}

\author{Angel Lozano\thanks{Angel Lozano (angel.lozano@upf.edu) is with Universitat Pompeu Fabra (UPF), 08018 Barcelona, Spain.
His work is supported by the European Project FET 265578 "HIATUS".},
Robert W. Heath Jr.\thanks{Robert W. Heath Jr. (rheath@ece.utexas.edu) is with The University of Texas at Austin, Austin, TX 78704-0240.
His work is supported by the Army Research Lab Grant W911NF-10-1-0420 and the Office of Naval Research Grant N000141010337.}
and Jeffrey G. Andrews\thanks{Jeffrey G. Andrews (jandrews@ece.utexas.edu) is with The University of Texas at Austin, Austin, TX 78704-0240  His work was supported by the National Science Foundation CIF-1016649.  Parts of this paper were presented at the 2012 Information Theory and Applications Workshop.}}

\begin{document}

\maketitle

\begin{abstract}
Cooperation is viewed as a key ingredient for interference management in wireless systems. This paper shows that cooperation has fundamental limitations. The main result is that even full cooperation between transmitters cannot in general change an interference-limited network to a noise-limited network. The key idea is that there exists a spectral efficiency upper bound that is independent of the transmit power. First, a spectral efficiency upper bound is established for systems that rely on pilot-assisted channel estimation; in this framework, cooperation is shown to be possible only within clusters of limited size, which are subject to out-of-cluster interference whose power scales with that of the in-cluster signals. Second, an upper bound is also shown to exist when cooperation is through noncoherent communication; thus, the spectral efficiency limitation is not a by-product of the reliance on pilot-assisted channel estimation. Consequently, existing literature that routinely assumes the high-power spectral efficiency scales with the log of the transmit power provides only a partial characterization. The complete characterization proposed in this paper subdivides the high-power regime into a \emph{degrees-of-freedom regime}, where the scaling with the log of the transmit power holds approximately, and a \emph{saturation regime}, where the spectral efficiency hits a ceiling that is independent of the power. Using a cellular system as an example, it is demonstrated that the spectral efficiency saturates at power levels of operational relevance.
\end{abstract}

\section{Introduction}

Wireless networks with many uncoordinated transmitters and receivers using the same spectrum are interference-limited, which means that increasing the transmit power of each node does not improve the spectral efficiency once the power is sufficiently high.  In cellular systems, this results in a large fraction of users having low signal-to-interference-and-noise ratio (SINR) regardless of the transmit powers, unless bandwidth-wasting methods such as frequency reuse are implemented to relieve edge users.  Similar effects can be observed in other networks: in WiFi, inefficient contention-based MAC protocols protect receivers from interference by silencing nearby transmitters.

It has been persuasively argued in a by-now vast literature (e.g. \cite{SenErk03,LanTse04,JanHed04}) that this limitation is not fundamental, but rather an artifact of each transmit-receiver pair communicating autonomously rather than cooperatively.  If the various nodes could cooperate, the logic goes, the corresponding interference channel could be converted to a (perhaps very large) broadcast channel---for the downlink---or multiple access channel---for the uplink---with all the transmitters (resp. receivers) jointly encoding (resp. decoding).
In the cellular context, it seems from this line of work that an arbitrary number of base stations (BSs) could cooperate to achieve enormous spectral efficiency gains over the lone-BS model \cite{FosKar06}, with the only limitation being the amount of coordination that is practical \cite{SimSom09}.

This is currently a problem of considerable theoretical and commercial interest.  Incomplete but significant cooperation between BSs has been extensively attempted by industry and is still ongoing, broadly under the current moniker of ``coordinated multipoint" (CoMP).  The bit rate and latency of the backhaul links have limited the benefits of such cooperation thus far, with disappointing gains typically not exceeding 30\% \cite{3GPPCoMP,GorokhovCTW,Irm11}.  In fact, one respected group has even experienced a net loss from cooperative techniques when the various over-the-air overheads are accounted for \cite{QCOM_ITA12}.  Are these widespread observations a byproduct of current technology limitations that could be overcome with better/more backhaul sharing (e.g., over dedicated fiberoptic control channels), improved feedback and overhead techniques, and/or better encoding and decoding methods?  Or is there a fundamental limitation lurking beneath the surface, one that is independent of the particular technology?

\subsection{Background and Status Quo}
\label{symptoms}

There are currently about 3 million macrocellular BSs worldwide and that number is expected to reach 50 million by around 2015, once small cells (pico and femtocells) are incorporated \cite{AndCla12}.
In a typical urban area, thousands to tens-of-thousands of BSs occupy the same spectrum.  Because of the ubiquity and commercial importance of cellular networks, plus the clear scope for cooperation in them, and in order to be concrete and clear, we focus the discussion around BS cooperation.  However, the subsequent models and main results of this paper apply to any set of transmitters and receivers and their scope is not limited to cellular networks.

Evaluating the performance of cellular networks is a complex task. Large-scale computer simulations are ultimately necessary to verify the performance of any specific technique, but they are hardly the best way to devise and probe ideas, build intuition, and glean insights.  Most designs in cellular systems are therefore incubated in much simpler settings that represent a fragment of a system
and only eventually are they transplanted and trialed at the level of an entire system.
A typical such controlled setting (at a given time epoch) is given by the following relationship.
\begin{relationship}
\label{rel1}
The observation at receiver $n$ is
\be
Y_n = \sum_{k=1}^K H_{nk} \sqrt{P} X_k + Z_n \qquad\qquad n=1,\ldots,N
\label{jogging}
\ee
where $K$ and $N$ are the number of transmitters and receivers, respectively,
$X_k$ is the signal generated by transmitter $k$, normalized such that $P$ is its power, and
$H_{nk}$ is the channel from transmitter $k$ to receiver $n$. Power differences at the $K$ transmitters can be simply absorbed into the channel coefficients or the noise variances. The term $Z_n$ is the noise at receiver $n$, typically white and Gaussian, with some normalized variance.
\end{relationship}

The signals and channel coefficients in Relationship \ref{rel1} can be scalars, or else properly dimensioned vectors and matrices to accommodate MIMO (multiple-input multiple-output) techniques. For the sake of exposition, unless otherwise stated, we henceforth consider single-antenna transmitters and receivers. Thus, the elements of Relationship \ref{rel1}  are scalars.

Relationship \ref{rel1} subsumes, in effect, most basic information-theoretic channel settings.
\begin{itemize}
\item Single-user channel if $K=N=1$. 
\item Multiaccess channel (MAC) if $K>1$ and $N=1$.
\item Broadcast channel (BC) if $K=1$ and $N>1$.
\item Interference channel (IC) if $K=N$ with $K,N>1$.
\end{itemize}
Combinations of these basic settings are also possible. Relationship \ref{rel1} has led to characterizations of the single-user, MAC and BC capacity, as well as asymptotic notions like the DMT (diversity-multiplexing tradeoff) \cite{divmul}, and practical techniques such as multiuser MIMO.

Relationship \ref{rel1} has been used to study the IC both generically and in the context of cellular systems.  Despite the fact that the capacity of the IC is yet to be determined, Relationship \ref{rel1} has  led to advances in the understanding of the IC.
\begin{itemize}
\item The definition of relevant quantities such as the number of DoF (degrees of freedom) \cite{Cadambe08}.
\item The development of centralized cooperative schemes such as Network MIMO \cite{karakayali2006}.
\item The genesis of distributed cooperative solutions such as IA (interference alignment) \cite{MMK08},
Max-SINR (maximum signal-to-interference-plus-noise ratio) \cite{Gomadam,PetHea:Algorithms-for-the-MIMO-Interference:09}  and other forms of cooperative interference management \cite{ShiLuoHe11}.
\end{itemize}

\begin{figure}
  \centering
 \includegraphics[width=0.6\linewidth]{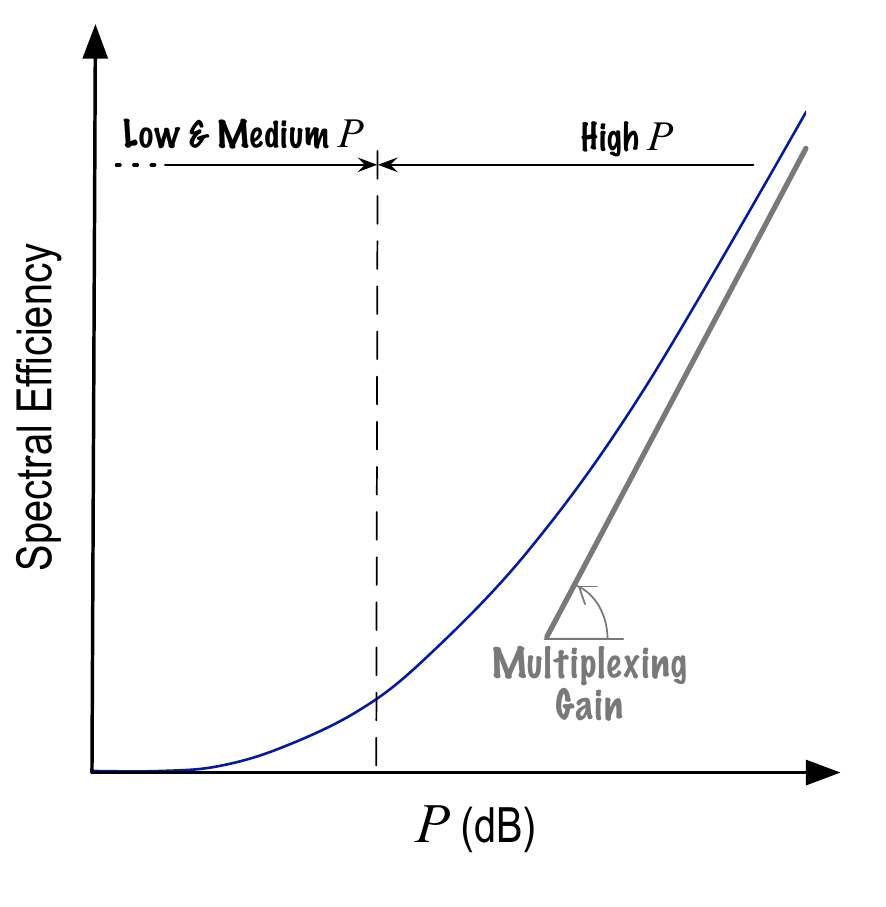}
  \caption{Spectral efficiency as function of $P$ (in dB) according to Relationship 1.}
  \label{DoF}
\end{figure}

Most development on the IC takes place in the high-power regime, which is where the nature of the IC
comes to the fore. For any setting conforming to Relationship \ref{rel1}, the high-power behavior is as illustrated in Fig. \ref{DoF}.
With the channel coefficients $\{ H_{nk} \}$ known by the corresponding receivers and possibly also the transmitters,
the spectral efficiency achieved by a given user grows, for $P \rightarrow \infty$, linearly with $\log (P)$ with a slope given by the so-called multiplexing gain. The multiplexing gain cannot exceed the number of DoF, and thus
cooperative techniques such as Network MIMO or IA aim at maximizing the number of DoF.

The IC developments spawned by Relationship \ref{rel1} have spurred an extensive amount of publications that invariably promise very large gains in spectral efficiency through cooperation (cf. \cite{GesHan10,WangWang11,SimeoneNOW12} and references therein).
However, as noted at the outset, in subsequent system-level simulations these gains do not seem to materialize: DoF maximizations attained through IA translate only to marginal gains or even outright losses in spectral efficiency \cite{tresch2009clustered,Mennerich11}. For example, \cite{SuhHoTse10}, showed that a $300\%$ gain from using IA in an isolated $3$-user setting shrinks to a mere $28\%$ gain when the same exact IA scheme is applied to a $19$-cell system.  (The conditions are otherwise identical: transmitters and receivers have $4$ antennas and the signal-to-noise ratio is $20$ dB.)  These serial many-fold discrepancies point to a disconnect, to some fundamental way in which a fragment of a cellular system is not properly modeled by Relationship \ref{rel1}.

\subsection{Modeling a Cluster Within a System}
\label{diagnosis}

An obvious problem with Relationship \ref{rel1} is that there is a cutoff of $K$ (possibly cooperating) transmitters, and all other interference is ignored.  The $K$ cooperating transmitters are typically geographical neighbors, and are referred to as a \emph{cluster}.  It is usually assumed that any interference from outside the cluster can be lumped into the $Z_n$ noise terms.  It cannot, because $Z_n$ has a fixed variance that does not depend on $P$ whereas the external interference power is proportional to $P$.  This can be illustrated by generalizing Relationship \ref{rel1} as follows, to more accurately describe a wireless network.

\begin{relationship}{\bf (Proposed System Model)}\\
\label{rel2}
The observation at receiver $n$ is
\be
Y_n = \sum_{k=1}^{K} H_{nk} \sqrt{P} X_k + \!\!\!  \sum_{k=K+1}^{\tilde{K}} \!\!\!\! H_{nk} \sqrt{P} X_k  + Z_n ~~ n=1,\ldots,N
\label{jogging2}
\ee
where $\tilde{K}$ and $\tilde{N}$ are the total numbers of transmitters and receivers in the system while $K$ and $N$ are the ones cooperating. Defining $Z_n' = \sum_{k=K+1}^{\tilde{K}} H_{nk} X_k$ as the out-of-cluster interference at receiver $n$,
\be
\! \; Y_n = \sum_{k=1}^{K} H_{nk} \sqrt{P} X_k + \sqrt{P} Z_n' + Z_n ~~\;\; n=1,\ldots,N
\label{jogging22}
\ee
\end{relationship}

\begin{figure}
  \centering
 \includegraphics[width=0.6\linewidth]{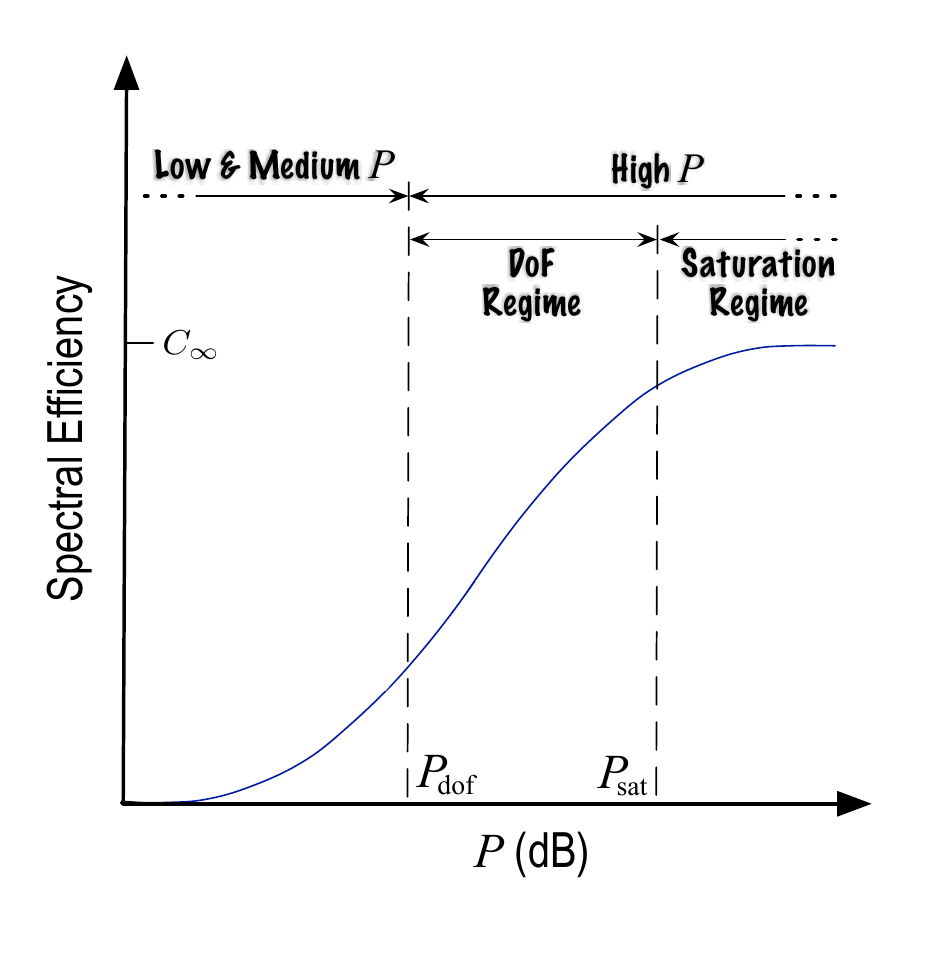}
  \caption{Spectral efficiency as function of $P$ (in dB) according to Relationship 2.}
  \label{fig:NewModel}
\end{figure}

Thus, as $P \to \infty$, the out-of-cluster interference does as well if the total network size is greater than $K$.   This gives rise to a \emph{saturation regime} as shown in Fig. \ref{fig:NewModel}, where further increasing the power above some value $\Psat$ does not noticeably increase the spectral efficiency because the external interference $Z'$ is larger than the noise $Z$.  Previous work had observed  that interference power scales with $P$ \cite{etkin2008gaussian,jafar2010generalized}, but the
representation therein differs from Relationship \ref{rel2} in two important ways.  First, in  \cite{etkin2008gaussian,jafar2010generalized}, the channel gains are deterministic and thereby known perfectly---at no cost---by all transmitters and receivers. In Relationship \ref{rel2}, in contrast, the channel gains outside the cluster of interest cannot be known (that is in fact what will come to define the boundaries of a cluster) and those within the cluster have to be learned, explicitly or not.  Second, in \cite{etkin2008gaussian,jafar2010generalized},  a specific ratio of noise to interference is required, whereas in Relationship \ref{rel2}  that can be arbitrary.  The assumptions in \cite{etkin2008gaussian,jafar2010generalized} leads to the conclusion that the spectral efficiency still scales indefinitely with $\log(P)$, only with a modified
DoF notion dubbed \emph{Generalized DoF}.  Relationship 2, in full generality, leads to a very different conclusion.

\subsection{Summary of Contributions}

This paper establishes that the correct representation of a cluster within a large wireless system is given by Relationship 2 and not Relationship \ref{rel1}, and that cooperation behaves very differently in the two models.  A possible objection to this claimed distinction could be that the high-power saturation only occurs because the cluster size $K$, i.e., the number of cooperating transmitters, is too small.  If $K$ were made large enough, wouldn't the saturation dissappear since interfering transmitters would be converted to cooperating ones?  The answer this paper establishes is negative.  Saturation of the spectral efficiency is unavoidable, and cooperation can at best push it into higher power levels.
This limitation holds in wireless networks with any degree of dynamics', i.e., channels that change over time or users that enter and leave the system.  Virtually all wireless networks exhibit such dynamics.  Although the derived limits may or may not be observed in practice today, they impose a hard ceiling on the gains from cooperative communication that can be achieved in the future.  In particular, information theoretic fundamentals lead to the conclusion that it is not possible to convert a sufficiently large interference network to a BC or MAC regardless of the amount of cooperation.

The points established descend from the model detailed in Section \ref{modeling}) and a fairly straightforward analysis of it for pilot-assisted channel estimation (in Section \ref{sec:coherent}) as well as a more fundamental analysis (in Section \ref{sec:noncoherent}).  The main result of the paper is Proposition 1, which establishes that there is an inescapable spectral efficiency upper bound  in large cooperative networks.  A large system (asymptotic) expression for this bound is given in Proposition 2.  The main observations from the analysis and accompanying examples are as follows.

\begin{enumerate}

\item In large systems, the existence of out-of-cluster interference (whose power scales with $P$) is inevitable. This holds
regardless of whether channel coefficients are explicitly estimated or not, and regardless of the size of the clusters within the system.

\item As a result of the out-of-custer interference, a cluster within a cellular system is described by Relationship 2 rather than Relationship \ref{rel1}. The performance of any cooperative technique then no longer resembles Fig. \ref{DoF} but rather Fig. \ref{fig:NewModel}.

\item For high $P$, two distinct regimes can be identified: a \emph{DoF regime} where the noise dominates over the out-of-cluster interference and a \emph{saturation regime} where the two become comparable and the spectral efficiency chokes. The notion of DoF is only meaningful in the DoF regime.

\item Two new quantities of interest emerge for the saturation regime. The value $\Psat$ is the point where the transition to saturation is said to occur. The resulting limiting spectral efficiency is denoted by $C_\infty$. These new quantities depend on the system topology, the channel propagation laws, and the degree of user mobility. In most cases, the transition to the saturation regime takes place well within the range of operational interest and thus studies conducted using Relationship \ref{rel1} rather than (\ref{jogging22}) are bound to be misleading in terms of system-level performance.

\item The saturation takes place both with pilot-assistedcommunication and with noncoherent communication. Therefore the saturation does not occur simply because of explicit channel estimation or because of pilot overhead in the face of finite coherence times.  Nor can improved channel estimation solve the problem.

\end{enumerate}

\section{Detailed System Model}
\label{modeling}

This section is devoted to refining Relationship 2 and to describing the models used to embody it in the remainder of the paper.

\subsection{Large-Scale Modeling}

Let $G_{nk}$ be the \emph{average} channel power gain (associated with distance decay, shadowing, building penetration losses, and antenna patterns) between transmitter $k$ and receiver $n$
and let us also define the normalized channel power gains
\be
g_{nk} = \frac{G_{nk}}{\sum_{\kappa=1}^K G_{n\kappa}}
\ee
such that, for every $n$,
\be
\label{supercopa}
\sum_{k=1}^K g_{nk} = 1 .
\ee
Thus, $g_{nk}$ signifies the share of receiver $n$'s signal power that corresponds to transmitter $k$ and the set $\{ g_{nk} \}$ completely and compactly characterizes the relative average gains among all the nodes in the network.  We term the set $\{ g_{nk} \}$ the \emph{geometry profile}.

We can further absorb the various normalizations for each receiver $n$ into a signal-to-noise ratio ($\SNR_n$) that scales with $P$
and a signal-to-(out-of-cluster)-interference ratio ($\SIR_n$) that does not scale with $P$. Specifically,
\begin{align}
\label{repas}
\SNR_n & = \frac{\sum_{k=1}^{K} G_{nk}P}{N_0 B} \\
\SIR_n & = \frac{\sum_{k=1}^{K} G_{nk}}{\sum_{k=K+1}^{\tilde{K}} G_{nk}}
\label{dimecres}
\end{align}
where $N_0$ is the noise spectral density and $B$ the bandwidth.

Using the definitions in (\ref{supercopa}), (\ref{repas}) and (\ref{dimecres}), we can rewrite (\ref{jogging22}) as
\be
y_n =  \sqrt{\SNR_n}  \sum_{k=1}^K \sqrt{g_{nk}}  h_{nk} x_k + \sqrt{\frac{\SNR_n}{\SIR_n}} z_n' + z_n \qquad\qquad n=1,\ldots,N
\label{jogging3}
\ee
where the noise terms $\{ z_n \}$, the out-of-cluster interference terms $\{ z_n' \}$, the signals $\{ x_k \}$, and the fading channel coefficients $\{ h_{nk} \}$
are all mutually independent random variables normalized to be unit-variance.

If both the noise and the out-of-cluster interference are Gaussian, then (\ref{jogging3}) becomes
\be
y_n = \sqrt{\SNR_n}  \sum_{k=1}^K \sqrt{g_{nk}}  h_{nk} x_k + \sqrt{1+\frac{\SNR_n}{\SIR_n}} z_n''  \qquad\qquad n=1,\ldots,N
\label{jogging4}
\ee
where $z_n''$ is the aggregate noise-plus-interference, also Gaussian and unit variance, i.e., $z_n'' \sim \mathcal{N}_{\mathbb{C}}(0,1)$.
Since the Gaussian distribution correctly models thermal noise, and the out-of-cluster interference is made up of a large number of independent terms and
thus its distribution tends to be approximately Gaussian too, we
focus on (\ref{jogging4}) with $z_n'' \sim \mathcal{N}_{\mathbb{C}}(0,1)$.
Nevertheless, all of the points made henceforth apply (qualitatively) in the wider generality of (\ref{jogging3}), with $z_n$ and $z_n'$ having
different distributions.
Eq. (\ref{jogging4}) can be further rewritten into the following.

\begin{relationship}
\label{rel3}
The observation at receiver $n$ is given by
\begin{align}
\label{jogging5}
y_n & = \sqrt{\frac{\SNR_n \, \SIR_n}{\SNR_n + \SIR_n}} \sum_{k=1}^K \sqrt{g_{nk}}  h_{nk} x_k + z_n''  \\
& = \sqrt{\SINR_n}  \sum_{k=1}^K \sqrt{g_{nk}} h_{nk} x_k + z_n''  \qquad\qquad\qquad n=1,\ldots,N
\label{jogging6}
\end{align}
where $\SINR_n$ is the signal-to-interference-plus-noise ratio at receiver $n$, which equals the harmonic mean of $\SNR_n$ and $\SIR_n$, i.e.,
\be
\frac{1}{\SINR_n} = \frac{1}{\SNR_n} + \frac{1}{\SIR_n}.
\ee

\end{relationship}

For small $P$, then, $\SINR_n \approx \SNR_n$ whereas  for $P \rightarrow \infty$, $\SINR_n \rightarrow \SIR_n$.
The formulation on the basis of $\SNR_n$ and $\SIR_n$ is very general in that it captures not only the scaling with $P$, but also with other parameters such as cell size or noise variance.

\subsection{Small-Scale Modeling}

Referring back to Relationship \ref{rel3}, the small-scale fading is modeled as Rayleigh and thus the normalized fading coefficients satisfy $h_{nk} \sim \mathcal{N}_{\mathbb{C}}(0,1)$. We consider frequency-selective fading with coherence bandwidth $\Bc$ and thus the value of each $h_{nk}$ varies from subband to subband in an IID (independent identically distributed) fashion.
In terms of the temporal fading dynamics, both block- and continuous-fading are accommodated.

\begin{itemize}
\item With block fading, the channels within each subband hold constant for some time and then change to a different value also in an IID fashion.
If we denote by $\Tc$ the channel coherence in time, then---irrespective of how the signaling is arranged along the
time and frequency dimensions---the number of symbols over which the channel remains coherent is roughly $L= \Bc \Tc$.
\item With continuous fading,
the channels within each subband are discrete-time stationary and ergodic random processes with a Doppler spectrum $S_h(\cdot)$ that is bandlimited, i.e.,
\be
\label{lovells}
\left\{ \begin{array}{ll}
          S_h(\nu)>0 & \quad |\nu|\leq \fd \\
          S_h(\nu)=0 & \quad |\nu|> \fd
        \end{array}
        \right.
\ee
for some maximum Doppler frequency $\fd \leq 1/2$.
In cellular channels, $\fd = v / (\lambda \Bc)$, where $v$ is the velocity and $\lambda$ is the carrier wavelength.
\end{itemize}

Block fading is a coarse but effective approximation to continuous fading, and remarkable equivalences between the two have been uncovered \cite{JindalLozano09,LozPorr11}.
In particular, both models have been shown to be equivalent in terms of channel estimation MMSE (minimum mean-square error) for the case of a rectangular Doppler spectrum,
$S_h(\nu)= \frac{1}{2 \fd}$ for $|\nu|\leq \fd$.
With such a spectrum, the MMSE is equivalent to that of a block-fading channel with $L=\frac{1}{2 \fd}$
(cf. Section \ref{sec:coherent}).

\begin{example}
With the typical cellular values $\lambda=0.15$ m and $\Bc=370$ KHz, $\fd \approx 2.5 \times 10^{-5}$ for pedestrian velocities and $\fd \approx 5 \times 10^{-4}$ for vehicular velocities.
This maps to $L \approx 20,000$ and $L \approx 1000$, respectively.
\end{example}




\subsection{Exemplary Cellular System}

To close this section, let us introduce an exemplary system that shall be utilized throughout.

\begin{figure}
  \centering
 \includegraphics[width=0.7\linewidth]{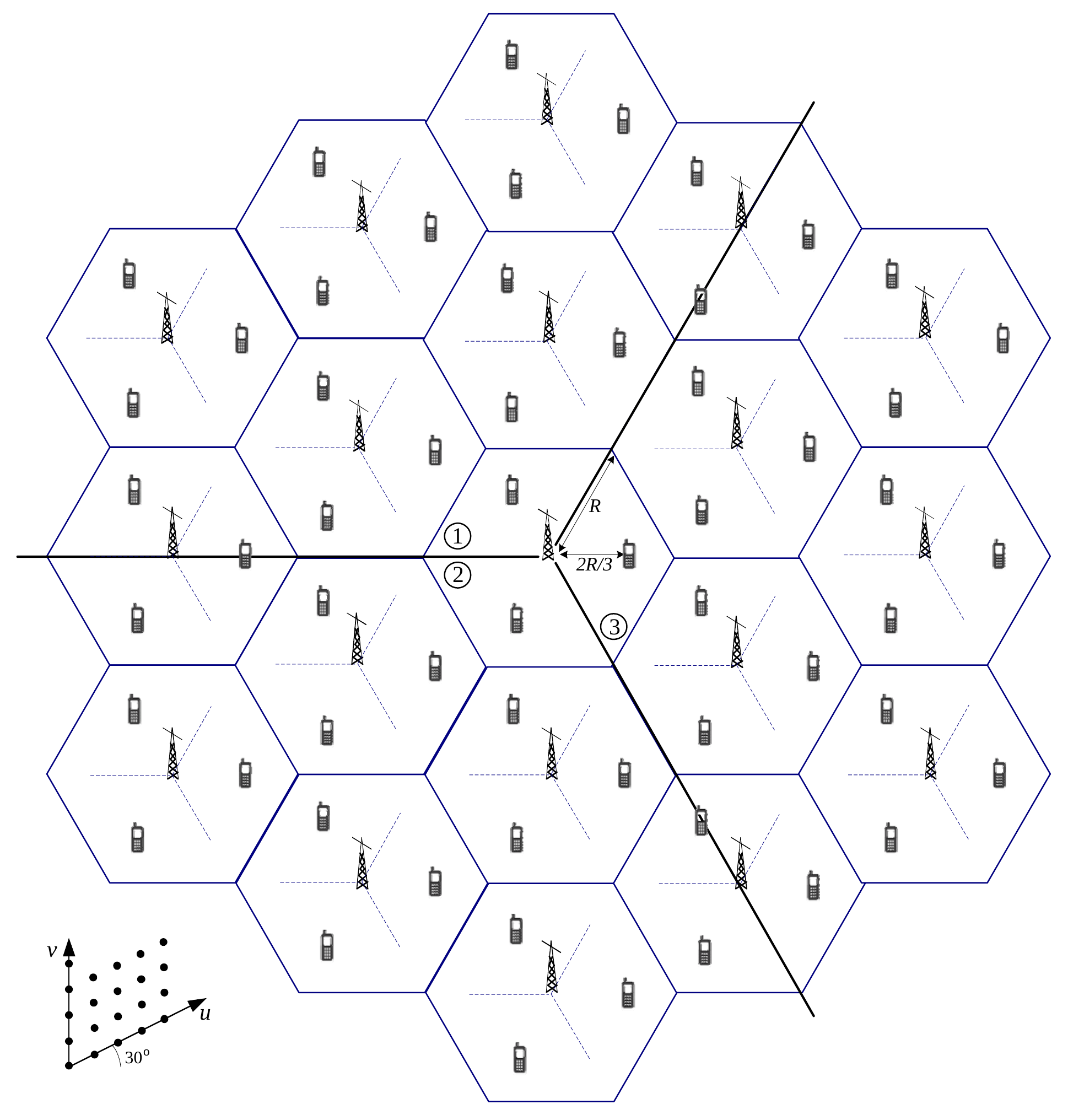}
  \caption{Regular hexagonal universe with tri-sector hexagonal cells of size $R$.
  Each user is centered within its sector and at distance $2R/3$ from the BS.}
  \label{hexuniverse}
\end{figure}

\begin{example}
\label{exemplary}
Consider a cellular system with tri-sector hexagonal cells of size $R$.
Each sector's antenna has a uniform gain over the $120^{\circ}$ span of the sector and a $Q|_{\rm dB}$-lower uniform gain outside that span.
Orthogonal signaling resources (time slots and frequency subcarriers) are allocated to the users within each sector. On any given resource, thus,
there is a single user per sector and hence $K=N$.
Each user is centered in azimuth within its sector and at distance $2R/3$ from the BS.
Depicted in Fig. \ref{hexuniverse} are an arbitrary reference cell and the two tiers around it, yet the system has infinitely many cells.
The signals experience distance-dependent decay with an exponent $\gamma$ and also Rayleigh fading.
\end{example}

Example \ref{exemplary} is representative of a cellular system, while having the virtue of being isotropic and of having regular user
locations. Eq. (\ref{jogging4}) can be embodied for any desired values of $K=N$, i.e., for clusters of arbitrary size, and the out-of-cluster interference can be easily
summed for $\tilde{K},\tilde{N} \rightarrow \infty$ thanks to the regular user locations.

\section{Pilot-Based Channel Estimation}
\label{sec:coherent}

In this section we establish that the spectral efficiency inevitably saturates for large $P$, under the restriction that the receivers estimate the channel coefficients on the basis of pilot symbols. We shall remove this restriction in the sequel.   The rationale for studying systems featuring explicit channel estimation followed by coherent detection of payload data \cite{hassibi} is that
virtually every existing wireless system abides by this procedure.
References on cooperation with pilot-based channel estimation include
\cite{venkatesan2007network,venkatesan2009wimax,tresch2009clustered,huang2009increasing,ramprashad2009cellular,PetersHeathPartitioning}.

The procedure entails the transmission of pilot symbols regularly in time (and in frequency when multiples of $\Bc$ are spanned).
We note that
(\emph{i}) pilot symbols are overhead, and
(\emph{ii}) the periodicity of pilot transmission is determined by the channel coherence.
Only a finite number of pilot symbols can be transmitted within a given coherence interval 
containing  $L$ symbols.
At least one pilot symbol, and most likely multiple ones, must be devoted to enabling the estimation of each channel coefficient at the corresponding receiver for each coherence
interval. If the channels were matrix-valued, as in the MIMO case, separate pilot symbols would be needed for every entry therein.
The number of channel coefficients that can
be estimated is therefore limited by $L$ and must be substantially lower than $L$ for the overhead not to be overwhelming.
Explicit channel estimation can thus occur only within clusters of limited dimension, inevitably with out-of-cluster interference from all the transmitters
beyond. Excessively large clusters incur excessive overhead and/or poor channel estimates that ultimately nullify the benefits of cooperation.

To make this intuition specific, denote by $\alpha$ the share of the symbols reserved for pilots; the rest, $(1-\alpha)$, is for payload data.
The pilot transmissions should be orthogonally multiplexed from each of the transmitters \cite{marzetta} and thus
the estimation of each of the $K$ channel coefficients at each receiver relies on a share $\alpha/K$ of the symbols.
When only the $k$th transmitter is actively sending pilots, the SINR at receiver $n$ is $g_{nk} \SINR_n$. With block fading, the MMSE on the estimation of $h_{nk}$ is then \cite{hassibi}
\be
\label{awards}
\MMSE_{nk} = \frac{1}{1+ g_{nk} \SINR_{n}  L \alpha/K}
\ee
whereas, with continuous fading \cite{JindalLozano09}
\be
\label{cttc}
\MMSE_{nk} = 1 - \int_{-\fd}^{\fd} \frac{ g_{nk} \SINR_{n}  S^2_h(\nu) }
{K/\alpha + g_{nk} \SINR_{n}  S_h(\nu)} d\nu.
\ee
If the Doppler spectrum $S_h(\cdot)$ is rectangular, (\ref{cttc}) becomes
\be
\label{upm}
\MMSE_{nk} = \frac{1}{1+ g_{nk} \SINR_{n}  \frac{\alpha/K}{2 \fd}}
\ee
which coincides with (\ref{awards}) for $\fd = \frac{1}{2L}$.
The MMSE expression in (\ref{upm}) thus allows embracing both the block- and continuous fading models in a single framework.

Denoting the estimate by $\hat{h}_{nk}$ and the estimation error by $\tilde{h}_{nk}$, it follows that $h_{nk} = \hat{h}_{nk} + \tilde{h}_{nk}$ with
$\hat{h}_{nk}$ and $\tilde{h}_{nk}$ uncorrelated and with $\E[ | \tilde{h}_{nk} |^2]=\MMSE_{nk}$. Hence, from Relationship \ref{rel3},
\be
\label{valencia}
y_n = \sqrt{\SINR_n}  \sum_{k=1}^K \sqrt{g_{nk}}  \hat{h}_{nk} x_k + \sqrt{\SINR_n}  \sum_{k=1}^K  \sqrt{g_{nk}}  \tilde{h}_{nk} x_k  + z_n''  \qquad\qquad n=1,\ldots,N.
\ee
Typically the receivers apply minimum-distance decoding utilizing the channel estimates as if they were correct, in which case the terms
in the second summation in (\ref{valencia}) play the role of additional Gaussian
noise \cite{amos-shamai}. With that, the effective SINR at receiver $n$ upon payload data detection is
\be
\SINR^{{\sf eff}}_{n} = \frac{\SINR_n \sum_{k=1}^K g_{nk} (1-\MMSE_{nk})}{1+\SINR_n \sum_{k=1}^K g_{nk} \MMSE_{nk}}
\ee
and the average spectral efficiency (bits/s/Hz/user) that can be attained reliably is
\be
C = (1-\alpha) \frac{1}{N} f \! \left( \SINR^{\sf eff}_1, \cdots, \SINR^{\sf eff}_N  \right)
\ee
where the function $f(\cdot)$ depends on the type of cooperation among the $K=N$ users.
The spectral efficiency, which vanishes for both $\alpha \rightarrow 0$ and $\alpha \rightarrow 1$,
is maximized by a proper choice of $0 < \alpha < 1$.

As $K$ grows, $\alpha$ needs to grow at least linearly with it to maintain the effective SINRs. Since enlarging the cooperation clusters
entails involving progressively weaker signals, $\alpha$ might have to grow superlinearly with $K$. Ultimately, $C$ is bound to peak at some cluster size
and diminish thereafter.

\begin{figure}
  \centering
 \includegraphics[width=0.8\linewidth]{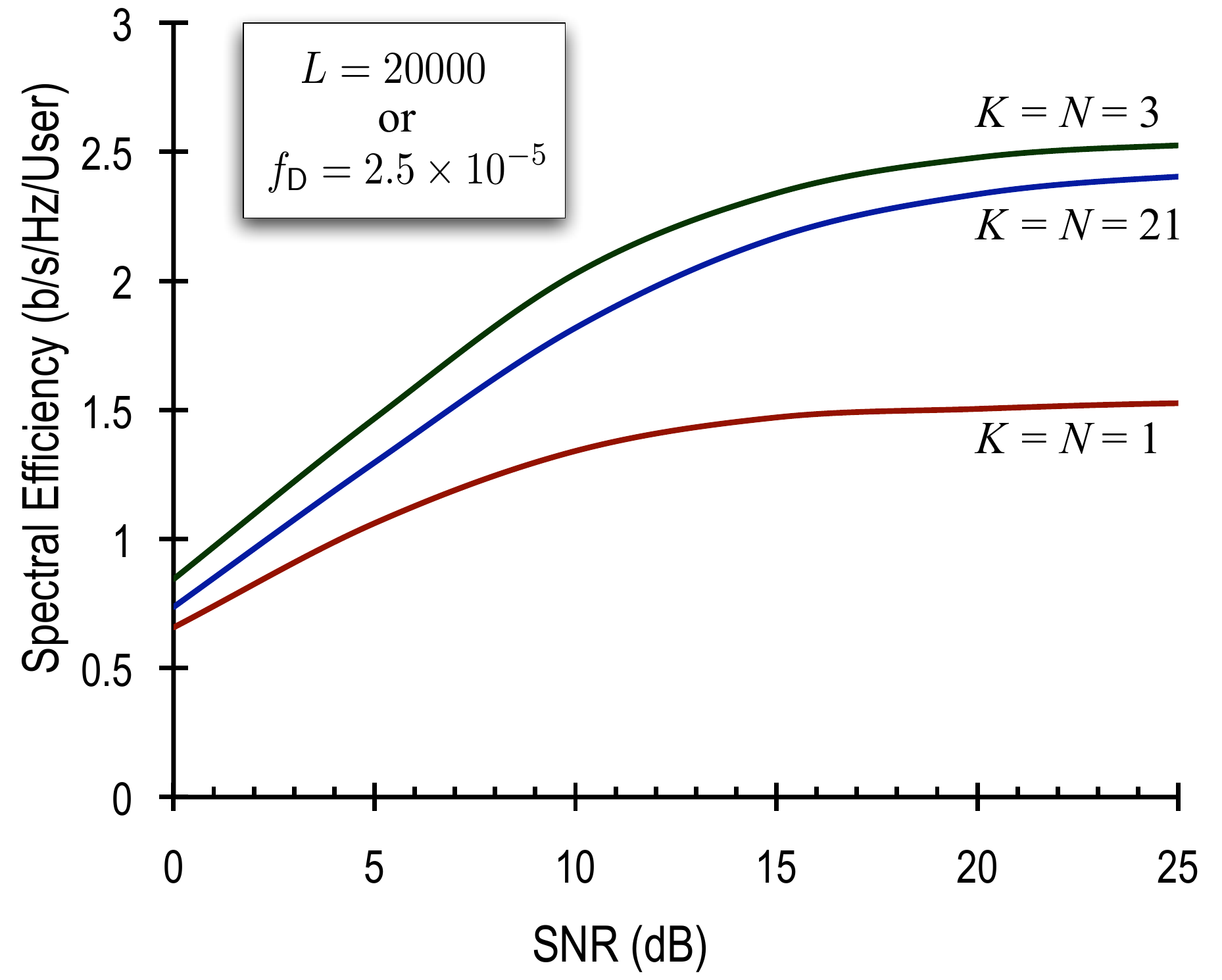}
  \caption{Spectral efficiency as function of $\SNR$ for varying cluster sizes with either block fading ($L=20,000$) or continuous fading ($\fd =2.5 \times 10^{-5}$).
  Uplink with full cooperation.}
  \label{ExplChEst}
\end{figure}

\begin{example}
\label{gladiator}
Consider the uplink of the system in Example \ref{exemplary} with clusters of arbitrary size $K=N$ where the BSs fully cooperate via Network MIMO, i.e.,
they jointly decode the $K$ signals received at the $N$ sectors.
The distance-decay exponent is $\gamma=3.8$ whereas $Q|_{\rm dB}=20$ dB.
Defining an $N \times K$ matrix $\Sm$ whose entries are independent and such that the $(n,k)$th entry is $\Sm_{nk} \sim \mathcal{N}_{\mathbb{C}}(0,\sigma^2_{nk})$ where
\be
\sigma^2_{nk} = \frac{g_{nk} \SINR_n (1-\MMSE_{nk})}{1+\SINR_n \sum_{\kappa=1}^K g_{n\kappa} \MMSE_{n\kappa} }  ,
\ee
the average spectral efficiency (bits/s/Hz/user) is
\be
C = (1-\alpha) \frac{1}{N} \E [ \log_2\det (\Idm + \Sm \Sm^\dagger) ] .
\ee
Note that, because of isotropy, $\SNR_n = \SNR$ for $n=1,\ldots,N$.
The fading is either block ($L=20,000$) or continuous with a rectangular Doppler spectrum ($\fd=2.5 \times 10^{-5}$), which are equivalent in terms of channel estimation
and correspond to pedestrian velocities.
Shown in Fig. \ref{ExplChEst} is the spectral efficiency for several cluster sizes. The baseline $K=N=1$ corresponds to single-user decoding, i.e., no cooperation.
The case $K=N=3$ corresponds to a cluster of $3$ facing sectors as in Fig. \ref{hexuniverse}.
Finally, the case $K=N=21$ corresponds to a cluster of $7$ cells: one central
cell plus the first tier around it. The pilot overhead $\alpha$ is equal for all users but otherwise it is optimized to maximize $C$, i.e., it is optimized for every cluster size and power level.
For details on how the out-of-cluster interference is computed, see Appendix \ref{example1}.
\end{example}

\begin{figure}
  \centering
 \includegraphics[width=0.8\linewidth]{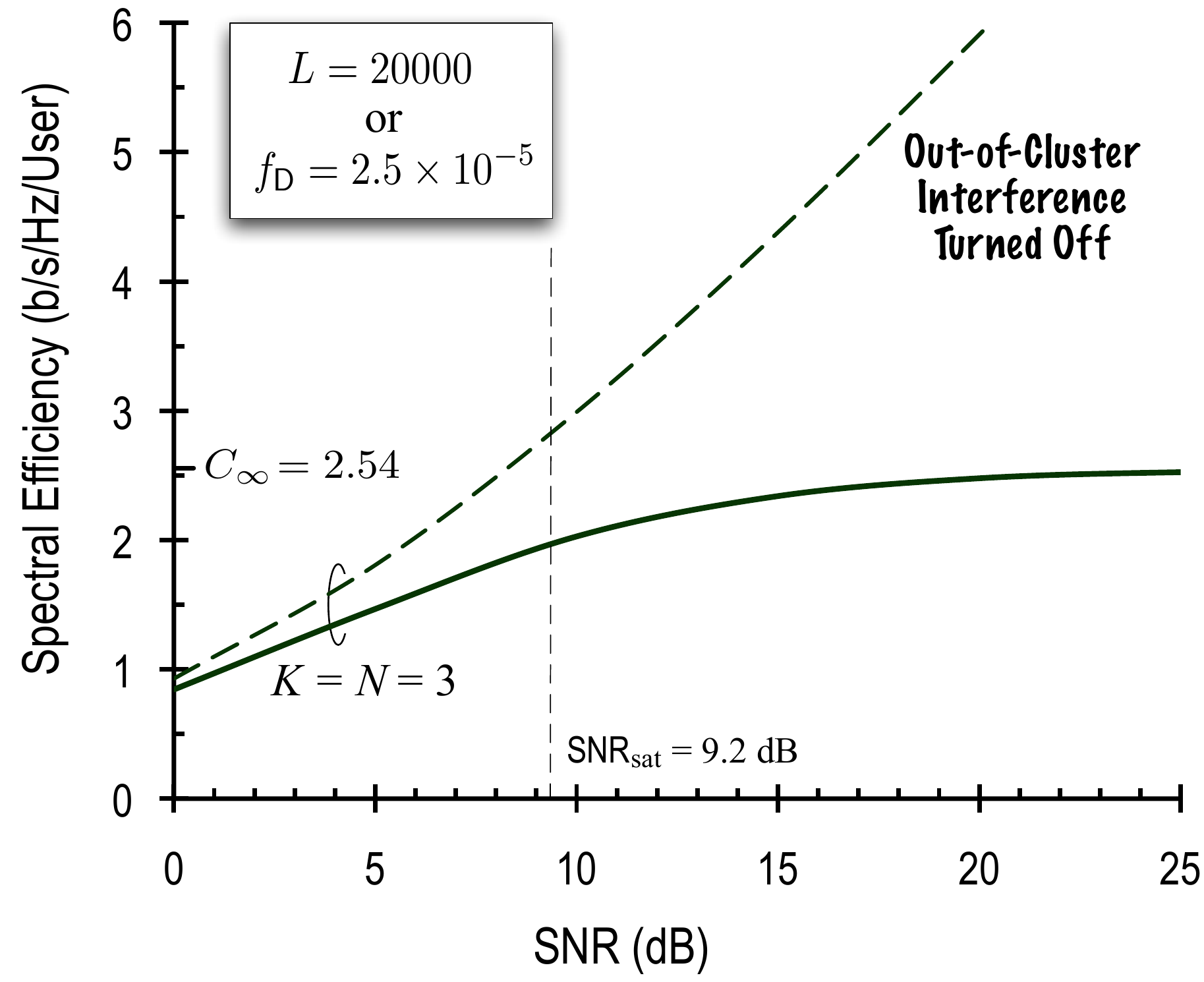}
  \caption{Spectral efficiency as function of $\SNR$ for $K=N=3$ with either block fading ($L=20,000$) or continuous fading ($\fd =2.5 \times 10^{-5}$).
  Uplink with full cooperation. In solid, with out-of-cluster interference included. In dashed, with out-of-cluster interference turned off.}
  \label{ExplChEst2}
\end{figure}

The performance in Example \ref{gladiator} improves when the cluster size goes from $1$ to $3$, but then degrades when the cluster size increases to $21$ and continues
to degrade for even larger sizes.
To gauge the impact of out-of-cluster interference, we re-plot the spectral efficiency corresponding to $K=N=3$ in
Fig. \ref{ExplChEst} next to that obtained in the same exact conditions only with all the transmitters outside the cluster turned off.
This comparison, presented in Fig. \ref{ExplChEst2}, evidences that without out-of-cluster interference we recover the traditional behavior (cf. Fig. \ref{DoF}).
Out-of-cluster interference, however, drastically modifies that behavior (cf. Fig. \ref{fig:NewModel}).
Note that modeling out-of-cluster interference as additional noise of fixed variance cannot fix the representation in Relationship \ref{rel1}, as it merely would shift the spectral
efficiency by some fixed amount.
Only the representation in Relationships \ref{rel2}--\ref{rel3} can properly reproduce the correct behavior.

\begin{example}
\label{pujol}
For Example \ref{gladiator} with $K=N=3$, the SIR that should be inserted into Relationship \ref{rel3} can be found to be
\begin{align}
\label{oriol}
\SIR_n & = \frac{\sum_{k=1}^3 G_{nk}}{\sum_{k=4}^{\infty} G_{nk}} \\
& = 9.2 \; \mathrm{dB} \qquad\qquad n=1,2,3
\end{align}
which, indeed, corresponds with the inflection point observed in Fig. \ref{ExplChEst2}.
That point, which corresponds to $\Psat$ and thus is denoted by $\SNRsat$, delineates the transition between the DoF
and the saturation regimes. In turn, the limiting spectral efficiency can be found to be $C_\infty = 2.54$ bits/s/Hz/user.
\end{example}

For user locations different from the regular ones in Example \ref{gladiator}, the values of $\{ \SIR_n \}$ vary and thus the transition between the
DoF and the saturation regimes takes place at different points, but qualitatively speaking the behavior is unaltered.

\begin{figure}
  \centering
 \includegraphics[width=0.8\linewidth]{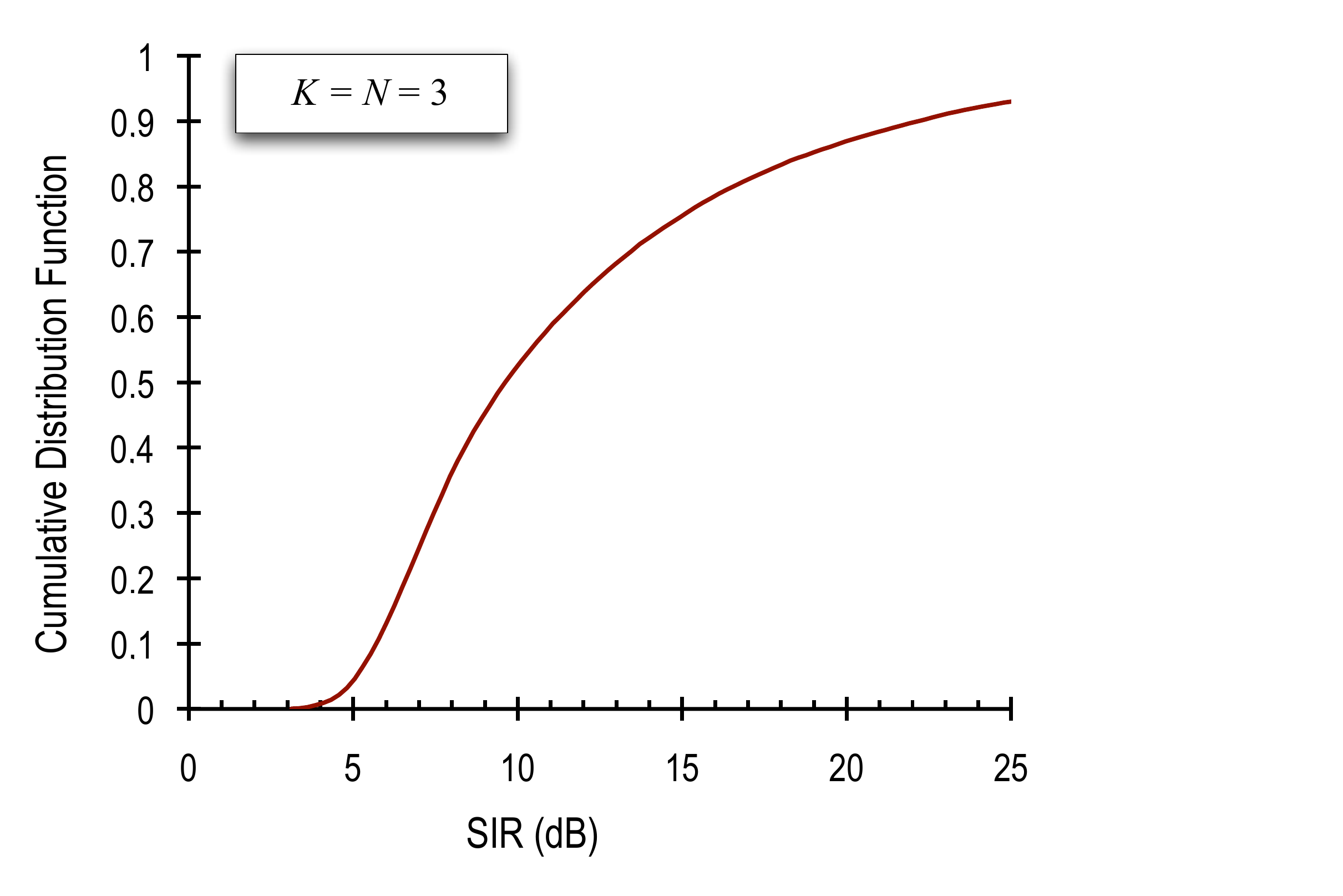}
  \caption{Cumulative distribution of SIR (dB) over all in-cluster locations for $K=N=3$. Uplink with $K=N=3$ and full cooperation.}
  \label{SIR3}
\end{figure}

\begin{example}
Consider a variation of Example \ref{gladiator}, with the out-of-cluster users still centered within their own sectors but with the location of the in-cluster users randomized.
Shown in Fig. \ref{SIR3} is the cumulative distribution of SIR at each of the three in-cluster receivers.
In almost $90\%$ of locations, the SIR is below $20$ dB. The median SIR is at $9.6$ dB, confirming that the value obtained in Example \ref{pujol} is indeed quite representative of the average conditions.
\end{example}

The figures derived in the preceding example indicate that, in a vast majority of cases, the spectral efficiency saturates at SNR levels of operational interest.
The final steps to fully generalize these figures would be to randomize the location of the out-of-cluster users and to incorporate shadow fading on all the links.
Exactly summing the out-of-cluster interference becomes challenging in that broad generality, but the computation is otherwise conceptually identical and the resulting SIR distribution
is not expected to depart much from the one in Fig. \ref{SIR3}.

Qualitatively similar observations to the ones made in this section for the uplink
can be made for the downlink, which could be analyzed by extending the pilot-assisted scheme
in \cite{caire2010multiuser} to the multicell realm.

\section{Noncoherent Detection}
\label{sec:noncoherent}

In this section we consider noncoherent detection without explicit channel estimation. From an information-theoretic perspective, noncoherent detection subsumes as a special case the procedure of transmitting pilot symbols, estimating the channel coefficients, and detecting the data coherently. Pilot symbols are a specific form of redundancy and explicit channel estimation followed by coherent data detection is a specific detection strategy.
It could be argued that  the need for clusters is not fundamental but simply a byproduct of having limited opportunities (due to the finite coherence) for pilot transmission.
If explicit channel estimation could be transcended by means of noncoherent detection, then perhaps the entire system could cooperate as one.
Following this argument, there would be no out-of-cluster interference and Relationship \ref{rel1} would remain a valid representation. In this section we refute this idea and show that, in large networks, the saturation of the spectral efficiency is still unavoidable and Relationship \ref{rel3} remains the correct representation for a cellular system or any cluster therein.

In the absence of CSI (channel-state information) at the receivers, the capacity-achieving signals and the capacity itself are generally unknown---even for a single-user channel.
The spectral efficiency achievable with complex Gaussian signals is also unknown, save for MIMO channels with IID entries \cite{Rusek12}.
Given the difficulty in computing the exact spectral efficiency achievable without CSI at the receivers, we instead show that its limiting value for $P \rightarrow \infty$, $C_\infty$,
is bounded by above by a quantity that does not depend on $P$; hence, $C$ cannot grow without bound with $P$.

With clusters of limited dimension, the presence of out-of-cluster interference is sure to bring about a finite $C_\infty$ with noncoherent detection just as it did with explicit channel estimation.
The question that lingers is whether this changes for $K=\tilde{K}$ and $N=\tilde{N}$, i.e., when the entire system cooperates as one.

Consider the uplink.
With block fading, the transmit-receive relationship can be vectorized for the entire system and all the symbols in a fading block as
\be
\label{upvec}
\Ym =  \diag \left \{ \sqrt{\SNR_1}, \ldots, \sqrt{\SNR_N} \right \} \Hm \Xm + {\bf Z}
\ee
where $\Ym$ and ${\bf Z}$ are $N \times L$, $\Hm$ is $N \times K$ and $\Xm$ is $K \times L$.
The entries of ${\bf Z}$ are IID with ${\bf Z}_{n \ell} \sim \mathcal{N}_{\mathbb{C}}(0,1)$
whereas the entries of $\Hm$ are independent with $\Hm_{nk} \sim \mathcal{N}_{\mathbb{C}}(0,g_{nk})$;
notice that the normalized channel power gains $\{ g_{nk} \}$ are directly incorporated as the variances of the entries of $\Hm$.
The entries of $\Xm$ are unit variance and, since
the $k$th row of $\Xm$ contains the signal sequence transmitted by user $k$ over the $L$ symbols of a fading block, the rows of $\Xm$ are independent.
Since the entire system is represented in (\ref{upvec}), there is no out-of-cluster interference.

\begin{proposition}
\label{enano}

Consider the uplink of a cellular system subject to block-fading with $K > L$.
Define, for $n=1,\ldots,N$, respective diagonal matrices
$
\Gm_n = \diag \{ g_{n1},\ldots,g_{nK} \}
$.
Each $\Gm_n$ contains along its diagonal the normalized power gains between the $K$ transmitters and the $n$th receiver and thus $\Tr \{ \Gm_n \}=1$.
If $\Xm$ is full rank, the average spectral efficiency that can be achieved reliably for $P \rightarrow \infty$ satisfies $C_\infty \leq C_\infty^{\sf \scriptscriptstyle UB}$ with
\be
\label{estiu}
C_\infty^{\sf \scriptscriptstyle UB} = - \frac{1}{K}  \sum_{n=1}^N \frac{1}{L} \, \E \left[  \log_2 \det \left( \Xm^\dagger \Gm_n \Xm \right) \right ] .
\ee

{\bf Proof:} See Appendix \ref{upbounduplink}.
\end{proposition}

In light of (\ref{supercopa}), $C_\infty^{\sf \scriptscriptstyle UB}$ remains bounded for $K,N \rightarrow \infty$.
In sufficiently large systems, therefore, there is no hope for $C_\infty$ to grow without bound with $P$.

Note also that the rank constraint on $\Xm$ is very mild, accommodating every signaling strategy utilized in wireless communication.
In particular, if the signal sequence transmitted by each user is IID complex Gaussian, then $C_\infty^{\sf \scriptscriptstyle UB}$ in (\ref{estiu}) can
be expressed using \cite[Prop. 4]{poweroffset} in a form that is closed although not particularly convenient to work with when $K$ and $L$ are large.

A more compact analytical handle on $C_\infty^{\sf \scriptscriptstyle UB}$ can be obtained by resorting to large-dimensional results in random matrix theory,
with the added advantage that
the expressions then hold for IID signals regardless of the distribution from which the symbols are drawn (e.g., PSK or QAM in addition to complex Gaussian).

\begin{proposition}
\label{bacteri}

Consider the uplink of a cellular system subject to block-fading with $K > L$.
Then,
\be
\label{torre}
C_\infty^{\sf \scriptscriptstyle UB}
\approx \frac{N}{K}  \log_2 e + \frac{1}{K} \sum_{n=1}^N \left[ \log_2 \frac{a_n}{L}  - \frac{1}{L} \sum_{k=1}^K \log_2 (1 + a_n g_{nk}) \right]
\ee
with each $a_{n}$ the nonnegative solution to
\be
\label{dembarra}
\sum_{k=1}^K \frac{g_{nk}}{g_{nk} + 1/a_n} = L
\ee
and with (\ref{torre}) becoming exact as $K, L \rightarrow \infty$.

{\bf Proof:} See Appendix \ref{marato}.

\end{proposition}

In certain special cases such as Example \ref{exemplary}, Propositions \ref{enano} and \ref{bacteri} simplify significantly.

\begin{corollary}
\label{marquet}
If the system is isotropic in the sense that the set of channel power gains from the transmitters looks the same from the vantage of each receiver,
i.e., every set $\{ g_{nk} \}$ for $n=1,\ldots,N$ can be reordered into a common set $ \{ g_k \}$, then we can define a unique matrix $\Gm$ containing such set on its diagonal
and rewrite (\ref{estiu}) as
\be
\label{estiu2}
C_\infty^{\sf \scriptscriptstyle UB} = - \frac{N}{K}  \frac{1}{L} \, \E \left[  \log_2 \det \left( \Xm^\dagger \Gm \Xm \right) \right ] .
\ee
In turn, Proposition \ref{bacteri} then specializes to
\be
\label{homo}
C_\infty^{\sf \scriptscriptstyle UB} \approx \frac{N}{K} \left[  \log_2 e +  \log_2 \frac{a}{L}  - \frac{1}{L} \sum_{k=1}^K \log_2 (1 + a g_{k})  \right]
\ee
with the unique $a$ being the nonnegative solution to
\be
\sum_{k=1}^K \frac{g_{k}}{g_{k} + 1/a} = L .
\ee
\end{corollary}

In the context of continuous fading channels, the expressions in both Proposition \ref{bacteri} and Corollary \ref{marquet} can be interpreted
using the equivalence $L = \frac{1}{2 \fd}$ put forth earlier in the paper.

Before applying Corollary \ref{marquet} to the infinitely large system in Example \ref{exemplary}, we verify the accuracy of the asymptotic approximation in (\ref{homo}) for
finite $K=N$ and for $L$ such that its non-asymptotic counterpart in (\ref{estiu2}) can be computed numerically.

\begin{example}
\label{nadal}
Consider the uplink of a square fragment of the system in Example \ref{exemplary} having $20 \times 20$ cells ($K=1200$) and let $L=100$.
All the sectors in the system cooperate fully via Network MIMO.
The distance-decay exponent is $\gamma=3.8$ whereas $Q|_{\rm dB}=20$ dB.
Monte-Carlo evaluation of (\ref{estiu2}) gives $C_\infty^{\sf \scriptscriptstyle UB} = 5.183$ b/s/Hz/user whereas
(\ref{homo}) gives $C_\infty^{\sf \scriptscriptstyle UB} = 5.181$ b/s/Hz/user.
\end{example}

Having verified its accuracy, let us apply Corollary \ref{marquet} to the full system in Example \ref{exemplary} and for values of $L$ in the range of practical interest.




\begin{example}
\label{almendrina}
Reconsider Example \ref{nadal} for $K=N \rightarrow \infty$.
For $L=20,000$, $C_\infty^{\sf \scriptscriptstyle UB}=11.86$ bits/s/Hz/user whereas, for $L=1000$, $C_\infty^{\sf \scriptscriptstyle UB} = 7.98$ bits/s/Hz/user.
For details, see Appendix \ref{example1}.
\end{example}

The spectral efficiencies actually achievable for $P \rightarrow \infty$ may be substantially lower than the values in Example \ref{almendrina},
not only because those correspond to upper bounds but also because the values of $L$ used for the computations, which are correct over short distances,
become optimistic in the context of cooperation across an entire system. The long propagation delays that arise when distant units cooperate is sure to lead to a longer delay
spread and thus a smaller $\Bc$, with the consequent reduction in $L$.

The exact saturation point notwithstanding, the spectral efficiency does saturate and thus Relationship \ref{rel1} cannot represent an entire system operating as one.
In contrast, Relationship \ref{rel3} does adequately model the saturation that occurs for large $P$.
A question that can be posed at this point is the following: if one wants to assume perfect CSI at the receivers, which values
for $\{ \SIR_n \}$ should be inserted into Relationship \ref{rel3}  to reproduce the results in Example \ref{almendrina}?
We can gauge this from the values of $C_\infty^{\sf \scriptscriptstyle UB}$ obtained therein.
Because of the isotropy in Example \ref{almendrina}, every user operates at the same spectral efficiency and thus $\SIR_n = \SIR$ for $n=1,\ldots,N$. With perfect CSI at the receivers, complex Gaussian signals are capacity-achieving and the uplink capacity of Relationship \ref{rel3} with full cooperation equals, for $P \rightarrow \infty$,
\be
C_\infty = \frac{1}{K} \log_2\det \left( \Idm + \SIR \, \Hm \Hm^\dagger \right) .
\ee
The structure of $\Hm$ satisfies the conditions of \cite[Thm. 5]{parisjournal} and thus, for $K=N \rightarrow \infty$,
\be
\label{tio}
C_\infty \stackrel{\scriptscriptstyle \rm a.s.}{\rightarrow} 2 \log_2 \left( \frac{1+\sqrt{1+4 \, \SIR}}{2} \right) - \frac{\log_2(e)}{4 \, \SIR} \left( \sqrt{1 + 4 \, \SIR} - 1 \right)^2 .
\ee
Solving for the $\SIR$ that equates (\ref{tio}) with the values for $C_\infty^{\sf \scriptscriptstyle UB}$ in Example \ref{almendrina} we obtain
\be
\begin{array}{ll}
          \SIR=39.96\,{\rm dB}, & L=20,000 \;({\rm pedestrian}) \\
          \SIR=28.02\,{\rm dB}, & L=1000 \;({\rm vehicular})
\end{array}
\ee
which, deriving from spectral efficiency upper bounds, are themselves upper bounds on the actual $\SIR$.
Although, with shadow fading and randomized user locations, these values are likely to vary significantly, they are indicative: even if an entire system were to be
operated as one with full cooperation, interference brings about a fundamental performance ceiling that corresponds to values of $\SIR$
within the range of interest in high-power analysis.
This ceiling, furthermore, depends exclusively on the coherence $L$, which relates to the degree of mobility, and the geometry profile $\{ g_{nk} \}$, which quantifies the degree of signal connectivity among users.

Besides computational convenience,
an additional benefit of the asymptotic approximations in Proposition \ref{bacteri} and Corollary \ref{marquet}
is that they cast light on how $C_\infty^{\sf \scriptscriptstyle UB}$ depends on the geometry profile.
\begin{itemize}
\item If $\{ g_{nk} \}$ is highly skewed for a given $n$, then most of the power received by BS $n$ corresponds to a few nearby users. Intuition then says that, given their relative strength, the fading of these users' channels and the overlaying
signals could be determined and most of the received power should be rendered useful. The asymptotic results confirm this
intuition: with fewer than $L$ nonnegligible terms, $C_\infty^{\sf \scriptscriptstyle UB}$ can be arbitrarily large and thus a sustained increase of the spectral efficiency with $P$ is feasible.
\item Alternatively, if $\{ g_{nk} \}$ for a given $n$
contains a myriad minute terms, rather than a few strong ones, each of these terms is simply too weak relative to the aggregate rest.
Intuitively, this should give rise to a bulk of residual interference that is fundamentally undecodable, and which is substantial in the aggregate even if each of the terms is by itself small. Again, the asymptotic results confirm this intuition: if $g_{nk} = 1/K$ for $k=1,\ldots,K$, then,
as $K$ grows without bound for fixed $L$, the interference becomes overwhelming and $C_\infty^{\sf \scriptscriptstyle UB}$ vanishes.
\end{itemize}

In actual systems, there will be a few strong signals in a sea of minute ones. The result is a finite value for $C_\infty^{\sf \scriptscriptstyle UB}$ that
depends on the skewness of the geometry profile. Note further that, since the $\{ g_{nk} \}$ are normalized, they are scale independent.
Cell size is therefore immaterial in terms of the geometry profile.

Note also that the schedulers that determine which user(s) in each cell are allocated to a given signaling resource play a significant role in establishing the geometry profile.
Subject to latency and quality-of-service constraints, the schedulers can therefore shape the geometry profile.
Dynamic definition of the cooperation clusters is also likely to be beneficial \cite{papadogiannis2008dynamic,garcia2010dynamic}.
Other aspects that may affect it include (possibly fractional) frequency reuse, power control, antenna patterns, and
antenna downtilting \cite{downtilt}.

For the downlink, the roles of transmission and reception and reversed relative to the uplink.
The corresponding noncoherent performance can be upper-bounded by allowing all the receivers to cooperate, in which case
the uplink derivations carry over and a performance ceiling is readily observed.
Tighter upper bounds might be obtained by removing the premise of receiver cooperation while considering downlink transmission strategies other than IID signaling.


\section{Concluding Discussion}
\label{remedy}

As argued up to this point, 
Relationship \ref{rel3} with the appropriate values for $\{ \SIR_n \}$ is the correct representation of a cellular system, or any fragment thereof.
The traditional high-power regime for user $n$, characterized by $\SNR_n \gg 1$, splits into two regimes (cf. Fig. \ref{fig:NewModel}):
\begin{enumerate}
\item The DoF regime where $\SNR_n \ll \SIR_n$. In this regime, the out-of-cluster interference is negligible relative to the noise and the spectral efficiency grows approximately linearly with $\log(P)$ according to the number of DoF computed without
out-of-cluster interference. The notion of DoF remains a valid approximation.
\item The saturation regime where $\SNR_n$ is comparable to or greater than $\SIR_n$. In this regime, the spectral efficiency chokes as it approaches $C_\infty$. 
The notion of DoF becomes meaningless; more precisely, the number of DoF is revealed to be zero.
\end{enumerate}

The transition between these regimes takes place, for user $n$, at $\SNR_{\mathsf{sat},n} \approx \SIR_n$. If the value of $\SIR_n$ is not explicitly known but $C_\infty$ is known,
the transition can alternatively be ascertained by assessing where an interference-free high-power expansion of the spectral efficiency intersects with $C_\infty$. It is pointless to operate much above $\SNRsat$.

\subsection{The Benefits of Cooperation}

We emphasize that the points made in this paper do not nullify the benefits of cooperation but, rather, they
show that cooperation has fundamental limitations that cannot be overcome through faster backhaul, more sophisticated signal processing, or any other technological advance.  Under Relationship \ref{rel3}, cooperation can still yield a markedly higher $C_\infty$ than if all interference was simply ignored.  Similarly, cooperation can provide an increased slope within the DoF regime\footnote{Recall that, although the number of DoF is strictly speaking zero, in this regime the slope is well approximated by the number of DoF with the out-of-cluster interference neglected}, allowing $C_\infty$ to be approached at lower power levels.  This is demonstrated in the following example.

\begin{figure}
  \centering
 \includegraphics[width=0.9\linewidth]{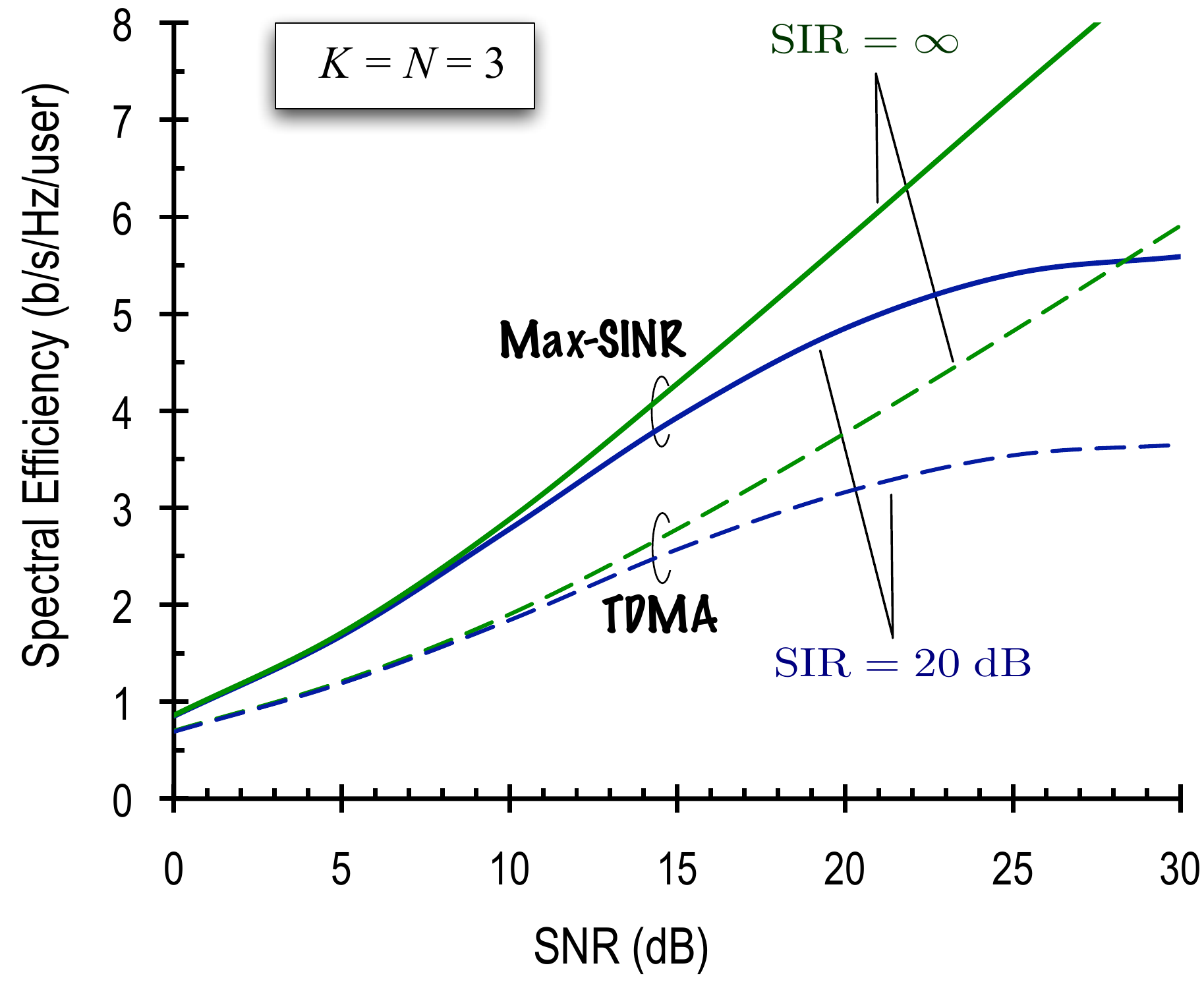}
  \caption{Max-SINR v. TDMA using Relationship \ref{rel3} with $K=N=3$ and with two antennas per transmitter and per receiver, for $\SIR = \infty$ and $\SIR = 20$ dB.}
  \label{ExampleNewModel}
\end{figure}

\begin{example}
\label{messibest}
Consider Relationship \ref{rel3} with $K=N=3$ and let every BS and user have two antennas.
Further let $\SNR_n=\SNR$, $n=1,2,3$.
Shown in Fig. \ref{ExampleNewModel} are the spectral efficiencies achieved by distributed Max-SINR \cite{Gomadam,PetHea:Algorithms-for-the-MIMO-Interference:09} and by round-robin TDMA for both $\SIR = \infty$ (no out-of-cluster interference)
and for $\SIR=20$ dB. Note that, for $\SIR = \infty$, Relationship \ref{rel3} reverts to Relationship \ref{rel1} and the theoretical number of DoF per user
($1$ with Max-SINR and $2/3$ with TDMA) are approached for $\SNR \rightarrow \infty$.
For $\SIR = 20$ dB, the inflection that delineates the DoF and saturation regimes occurs around $\SNR = 20$ dB as expected.
\end{example}

Apropos Example \ref{messibest} with $\SIR=20$ dB, we observe that the value of $C_\infty$ is over $50\%$ higher with Max-SINR relative to TDMA.  Further, in the DoF regime---from about 5 to 20 dB---a substantial difference between Max-SINR and TDMA builds up.
Altogether then, cooperation continues to provide a substantial advantage but the number of DoF is only a partial measure thereof.

\subsection{Future Directions}

This paper has shown that saturation of the spectral efficiency at sufficiently high powers is unavoidable in large systems.
A key insight is that interference is made up of a multitude of terms, the majority of which are too minute to be tracked.
Beyond a few strong terms, the structure of the interference is too intricate to discern yet the sum of all these minute terms is significant. Essentially, the receivers are near-sighted. They can only focus on a few strong nearby transmitters, and everything else in the distance looks fundamentally blurry. The lack of focus depends on the extent of mobility in the system.

Further work is needed to expand the observations in this paper. For example, the analysis focused on the uplink; the extension to the downlink is interesting and nontrivial. Only single-antenna transmitters and receivers were considered. It would be good to extend the results to MIMO, including large-dimensional MIMO systems that are limited by pilot contamination~\cite{Mar:Noncooperative-Cellular-Wireless:10}. In particular, it would be of interest to understand the tradeoffs between numbers of antennas at a single base station and coordination of many base stations to understand if there is any fundamental preference for centralized or distributed antenna architectures. If the system is small or the clusters are highly isolated (e.g., through geographical separation, penetration losses, or millimeter-wave propagation losses) then the $\SNRsat$ can be so large as to render the saturation anecdotal. It would be useful to understand and classify systems based on the proximity of $\SNRsat$ to practical operating SNRs say less than $30$dB.  The results in this paper neglected propagation delay among distant units. Characterizing the impact of delay (as a function of bandwidth) on $\SNRsat$ and $C_\infty$ would likely further reduce both of these values, making them more operationally relevant.  Finally, it would be good to know what an optimal cooperation cluster size is, in both the coherent and noncoherent realms.



\appendix

\subsection{Proof of Proposition \ref{enano}}
\label{upbounduplink}


Using the chain rule, the mutual information between $\Xm$ and $\Ym$ (in bits per coherence block) can be expressed as
\be
\label{buscantpis}
I(\Xm ; \Ym) = I(\Hm \Xm ; \Ym) - I(\Ym;\Hm | \Xm) .
\ee
Invoking (\ref{supercopa}) and the zero-mean unit-variance nature of the entries of $\Xm$, it can be verified that the entries of $\Hm \Xm$ are also zero-mean and unit-variance.
The term $I(\Hm \Xm ; \Ym)$ is then upper-bounded by the value it would take if those entries were
IID $\mathcal{N}_{\mathbb{C}}(0,1)$.
Thus,
\be
I(\Xm ; \Ym) \leq \sum_{n=1}^N L \log_2(1+\SNR_n) -  I(\Ym;\Hm | \Xm) .
\label{botswana}
\ee
Turning now our attention to $I(\Ym;\Hm | \Xm)$, and denoting differential entropy operator by $\mathfrak{h}(\cdot)$,
\be
\label{peke}
I(\Ym;\Hm | \Xm)=\mathfrak{h}(\Ym | \Xm) - \mathfrak{h}(\Ym | \Hm,\Xm).
\ee
Conditioned on $\Xm$, the rows of $\Ym$ are independent (both conditionally and unconditionally on $\Hm$)
and thus both differential entropy terms in (\ref{peke}) can be computed row-wise and simply added. It follows that
\be
I(\Ym;\Hm | \Xm) = \sum_{n=1}^N I({\bf y}_n;{\bf h}_n | \Xm)
\ee
where ${\bf y}_n = \sqrt{\SNR_n} \, {\bf h}_n \Xm + {\bf z}_n$ with ${\bf h}_n$ and ${\bf z}_n$ the $n$th rows of $\Hm$ and ${\bf Z}$, respectively.
Since ${\bf h}_n$ and ${\bf z}_n$ are complex Gaussian vectors,
\be
I(\Ym;\Hm | \Xm) = \sum_{n=1}^N \E \left[ \log_2 \det \left( \Idm +  \SNR_n \Xm^\dagger \Gm_n \Xm \right) \right]
\label{dollar}
\ee
where, recall, $\Gm_n = \diag\{ g_{n1},\ldots,g_{nK} \}$.

Since $I(\Xm ; \Ym)$ increases monotonically with $P$, we concentrate on upper-bounding it for $P \rightarrow \infty$.
Suppose first that we had $K < L$. Then, for large $\SNR_n$, $n=1,\ldots,N$, we would have
\be
I(\Ym;\Hm | \Xm) = \sum_{n=1}^N \left( K \log_2 \SNR_n + \E \left[ \log_2 \det \left( \Xm \Xm^\dagger \Gm_n \right) \right ]  \right) + o(1)
\ee
and, from (\ref{botswana}),
\be
I(\Xm ; \Ym) \leq  \sum_{n=1}^N \left(  (L-K) \log_2 \SNR_n - \E \left [ \log_2 \det \left(\Xm \Xm^\dagger \Gm_n \right) \right ] \right) + o(1)
\ee
whose right-hand side grows unboundedly with $P$ indicating that there is hope for $I(\Xm ; \Ym)$ to grow unboundedly with $P$. 

Conversely, when $K \geq L$, and given the full-rank condition of $\Xm$,
\be
I(\Ym;\Hm | \Xm) = \sum_{n=1}^N \left( L \log_2 \SNR_n + \E \left[ \log_2 \det \left( \Xm^\dagger \Gm_n \Xm \right) \right ]  \right) + o(1)
\ee
and thus (\ref{botswana}) becomes, for large $P$,
\be
I(\Xm ; \Ym) \leq - \sum_{n=1}^N \E \left [ \log_2 \det \left( \Xm^\dagger \Gm_n \Xm \right) \right ] + o(1)
\ee
from which the claimed upper bound in (\ref{estiu}) follows.

\subsection{Proof of Proposition \ref{bacteri}}
\label{marato}

If each user transmits an IID sequence, then the entries of $\Xm$ are IID and the asymptotic analysis in \cite{parisjournal,silverstein} can be applied to Proposition \ref{enano}.
Couched in the notation of this paper, we have that \cite[Section VI-A]{parisjournal}
\be
- \frac{1}{L}  \log_2 \det \left( \Xm^\dagger \Gm_n \Xm \right) \approx \log_2 e - \log_2 \left( \sum_{k=1}^K \frac{g_{nk}}{1 + a_{nk}} \right) - \frac{1}{L} \sum_{k=1}^K \log_2 (1 + a_{nk})
\ee
where the approximation becomes exact as $K,L \rightarrow \infty$, with $K \geq L$, and where
 $a_{nk}$ is the nonnegative solution to
\be
a_{nk} = \frac{L \, g_{nk}}{\sum_{j=1}^K \frac{g_{nj}}{1+ a_{nj}}}  .
\ee
Defining
\be
a_n = \frac{L}{\sum_{j=1}^K \frac{g_{nj}}{1+ a_{nj}}}
\ee
we can write $a_{nk} = a_n g_{nk}$ and
\be
- \frac{1}{L}  \log_2 \det \left( \Xm^\dagger \Gm_n \Xm \right) \approx \log_2 e + \log_2 \left( \frac{a_n}{L} \right) - \frac{1}{L} \sum_{k=1}^K \log_2 (1 + a_n g_{nk})
\ee
with $a_{n}$ the nonnegative solution to
\be
\sum_{k=1}^K \frac{g_{nk}}{g_{nk} + 1/a_n} = L .
\ee

\subsection{Details of Examples \ref{gladiator} and \ref{almendrina}}
\label{example1}

If the integers $u$ and $v$ denote the indices of a cell on the axes shown in the inset of Fig. \ref{hexuniverse}, the cartesian coordinates of the
corresponding BS are
\begin{align}
x & = \frac{3}{2} u R \\
y & = \sqrt{3} \left( v + \frac{u}{2} \right) R .
\end{align}
Relative to its serving BS, the relative cartesian position of a user centered in azimuth in its sector and at distance $2R/3$ from the base is,
for each of the sectors as labeled in Fig. \ref{hexuniverse},
\begin{align}
\Delta x_1 & = - \frac{R}{3} \qquad\quad \Delta x_2 = - \frac{R}{3} \qquad\quad \;\, \Delta x_3 = \frac{2R}{3} \\
\Delta y_1 & = \frac{R}{\sqrt{3}} \qquad\quad \; \Delta y_2 = - \frac{R}{\sqrt{3}} \qquad\quad \Delta y_3 = 0 .
\end{align}
Therefore, the distance between the BS at the origin ($u=v=0$) and the users at the cell with indices $u$ and $v$ equals, for each of the sectors therein,
\begin{align}
d_1(u,v) & = R \sqrt{\left( \frac{3}{2} u - \frac{1}{3} \right)^2 + 3 \left( v + \frac{u}{2} + \frac{1}{3} \right)^2} \\
d_2(u,v) & = R \sqrt{\left( \frac{3}{2} u - \frac{1}{3} \right)^2 + 3 \left( v + \frac{u}{2} - \frac{1}{3} \right)^2} \\
d_3(u,v) & = R \sqrt{\left( \frac{3}{2} u + \frac{2}{3} \right)^2 + 3 \left( v + \frac{u}{2} \right)^2} .
\end{align}
The corresponding normalized power gains are $g_1(u,v)=D d_1^{-\gamma}$, $g_2(u,v)=D d_2^{-\gamma} /Q$ and $g_3(u,v)=D d_3^{-\gamma} /Q$, with
$D$ the constant that renders
\be
\label{findD}
\sum_{u=-\infty}^\infty \sum_{v=-\infty}^\infty \left( g_1(u,v) + g_2(u,v) + g_3(u,v) \right) = 1
\ee
where $Q$ is the antenna front-to-back ratio. (Because of symmetries, applying the factor $1/Q$ to sectors $2$ and $3$
of each cell is equivalent to applying it to all three sectors in the slices of the system spanned by sectors $2$ and $3$ of the central reference cell.)
With $\gamma=3.8$, (\ref{findD}) gives $D=0.157 R$.
Then, for Example \ref{gladiator}, we can rewrite (\ref{dimecres}) as
\be
\label{abidal}
\SIR_n = \frac{\sum_{k=1}^{K} g_{nk}}{\sum_{k=K+1}^{\tilde{K}} g_{nk}}
\ee
and, using all the foregoing quantities, compute each $\SINR_n$ by having the appropriate terms in the numerator and denominator of (\ref{abidal}) with $\tilde{K} \rightarrow \infty$.

For Example \ref{almendrina}, $a$ is the solution to
\be
\label{country1}
\sum_{u=-\infty}^\infty \sum_{v=-\infty}^\infty \left( \frac{g_{1}(u,v)}{g_{1}(u,v) + 1/a}
+ \frac{ g_{2}(u,v)}{ g_{2}(u,v) + 1/a} + \frac{ g_{3}(u,v)}{ g_{3}(u,v) + 1/a}  \right)= L
\ee
while
\begin{align}
C_\infty = & \log_2 \frac{a e}{L}  - \frac{1}{L} \sum_{u=-\infty}^\infty \sum_{v=-\infty}^\infty \left[ \log_2 (1 + a \, g_{1}(u,v)) + \log_2 \left(1 + a  g_{2}(u,v) \right) \right. \non
& \left. + \log_2 \left(1 + a  g_{3}(u,v) \right) \right]
\label{country2}
\end{align}
Numerical evaluation of (\ref{country1}) and (\ref{country2}) with $Q=100$ yields the results in Example \ref{almendrina}.


\end{document}